\documentclass[oldversion, preprint]{aa}  
\usepackage{graphicx}
\usepackage{txfonts}
\usepackage{longtable}
\usepackage{multirow}
\begin{document}
   \title{Radio emission of SN1993J: the complete picture.}
   \subtitle{I. Re-analysis of all the available VLBI data.}

   \author{I. Mart\'i-Vidal\inst{1,2}
          \and
          J.M. Marcaide\inst{1}
          \and 
          A. Alberdi\inst{3}
          \and
          J.C. Guirado\inst{1}
          \and 
          M.A. P\'erez-Torres\inst{3}
          \and
          E. Ros\inst{2,1}
          }

   \institute{Dpt. Astronomia i Astrof\'isica, Universitat de Val\`encia,
              C/ Dr. Moliner 50, E-46100 Burjassot (Spain)\\
              \email{imartiv@mpifr-bonn.mpg.de}
         \and
             Max-Planck-Institut f\"ur Radioastronomie,
             Auf dem H\"ugel 69, D-53121 Bonn (Germany) 
         \and
             Instituto de Astrof\'isica de Andaluc\'ia (CSIC),
             C/ Camino bajo de Hu\'etor 50, E-18008 Granada (Spain)
             }

   \date{Accepted for publication in A\&A}
 
\abstract{
We have performed a complete re-calibration and re-analysis of all the available 
Very-Long-Baseline-Interferometry (VLBI) observations of supernova SN1993J, following an 
homogeneous and well-defined methodology. VLBI observations of SN1993J at 69 epochs, 
spanning 13 years, were performed by two teams, which used different 
strategies and analysis tools. The results obtained by each group are similar, but their conclusions 
on the supernova expansion and the shape and evolution of the emitting region differ significantly.
From our analysis of the combined set of observations, we have obtained an expansion curve with 
unprecedented time resolution and coverage. We find that the data from both teams are compatible 
when analyzed with the same methodology. One expansion index 
($m_3 = 0.87 \pm 0.02$) is enough to model the expansion observed at 1.7\,GHz, while two 
expansion indices ($m_1 = 0.933\pm0.010$ and $m_2 = 0.796\pm0.005$), separated by a break time, 
$t_{br} = 390\pm30$ days, are needed to model the data, at frequencies higher than 1.7\,GHz,
up to day $\sim4000$ after explosion. 
We thus confirm the wavelength dependence of the size of the emitting region reported by one
of the groups. We also find that all 
sizes measured at epochs later than day $\sim 4000$ after explosion are systematically smaller 
than our model predictions (i.e., an additional expansion index might be needed to properly 
model these data). We also estimate the fractional shell
width ($0.31 \pm 0.02$, average of all epochs and frequencies) and the level of opacity to the 
radio emission by the ejecta. We find evidence of a spectral-index radial gradient in the 
supernova shell, which is indicative of a frequency-dependent ejecta opacity. Finally, we study 
the distribution and evolution of the azimuthal anisotropies (hot spots) found around the radio 
shell during the expansion. These anisotropies have intensities of $\sim 20$\% of the mean flux 
density of the shell, and appear to systematically evolve during the expansion.   
}

\keywords{radio continuum: stars -- supernovae: general -- supernovae: individual: SN1993J 
          -- galaxies: individual: M81}
   \maketitle

\section{Introduction}
\label{I}

Supernova \object{SN\,1993J}, in the galaxy \object{M\,81},
has been one of the brightest supernovae ever in the radio band. The peak of 
emission at 5\,GHz was $\sim100$\,mJy (e.g. Weiler et al. \cite{Weiler2007}),
much larger than typical peak flux densities of radio supernovae. The large 
flux density of this supernova, together 
with its high declination, allowed for long observing 
campaigns with VLBI. Two research groups (one led by 
N. Bartel and the other one led by J.M. Marcaide) have monitored this supernova 
with the VLBI technique, from 1993 (Marcaide et al. \cite{Marcaide1994}, 
Bartel et al. \cite{Bartel1994}) to 2005. 

Different results on the structure and expansion of the radio-emitting region of 
SN\,1993J have been reported by both groups, based on the subset of VLBI data 
taken by each group. Marcaide et al. (\cite{Marcaide1997}) 
reported the first evidence of deceleration in the shell expansion (i.e., 
$R \propto t^m$, see Chevalier \cite{Chevalier1982a}), with an estimated expansion index 
of $m = $0.86$\pm$0.02. Bartel et al. (\cite{Bartel2002}) confirmed a deceleration, 
but claimed up to four changes in the value of 
$m$ corresponding to four different expansion periods and interpreted the changes in 
the expansion index as changes in the 
mass-loss wind of the progenitor star through the pre-supernova stage. 
However, from their set of VLBI observations, which range from day 182 to day 3867 
after explosion, Marcaide et al. (\cite{Marcaide2009}) found a wavelength-dependent expansion 
curve that can be modeled using only one expansion index ($m = 0.86$) for their 
low-frequency data data (1.7\,GHz) and two expansion 
indices (0.86 and 0.79, separated by a break time on day $\sim$1500 
after explosion) for the data at all higher frequencies. These 
authors interpret the frequency-dependent expansion curve as being
caused by (the possible combination of) two effects: 1) a changing (and frequency-dependent) 
opacity to the radio emission by the supernova ejecta\footnote{The ejecta are located 
behind the inner boundary of the radio shell and may block (partly or fully) the emission coming 
the backside of the shell; see Eq. B.1 of Marcaide et al. (\cite{Marcaide2009}) for a 
mathematical definition of the ejecta opacity in our shell model.} and 2) a radial drop in 
the amplified magnetic fields inside the radiating region combined with the finite 
sensitivity of the VLBI observations.

In this paper (Paper I), we report on a homogeneous analysis of the complete set of available 
VLBI observations of SN\,1993J (69 epochs), using different approaches to minimizing the effects 
of any possible bias in the data analysis. We studied the details of the expansion curve at 
several frequencies and the evolution of the structure of the radio shell throughout the 
history of the SN\,1993J radio emission, with unprecedented time resolution 
and coverage. We confirm earlier findings reported in Marcaide et al. 
(\cite{Marcaide2009}) and report a model of the expansion curve compatible
with the shell sizes obtained using different approaches. We also present a 
study of the distribution and evolution of inhomogeneities inside the shell. 
In another publication (Mart\'i-Vidal et al. \cite{PaperII}, hereafter Paper II), 
we present a new simulation code able to simultaneously model the expansion curve 
and the radio light curves of SN\,1993J reported by Weiler et al. (\cite{Weiler2007}). 
We then present the extensions to the Chevalier model 
(Chevalier \cite{Chevalier1982a}, \cite{Chevalier1982b}) to satisfactorily fit 
all the radio data. 

In Sect. \ref{II} we describe the complete set of VLBI observations of 
SN\,1993J, most of which we have re-analyzed {\em ab initio}. In Sect. \ref{III} 
we report on the location and proper motion of the radio shell. In Sect.
\ref{IV} we report on the complete expansion curve, obtained with different 
approaches, and in Sect. \ref{V} we discuss the departure in the evolution of 
the supernova structure from self-similarity.

\section{Observations}
\label{II}

In Table 1 
we show the complete set of available VLBI observations of SN\,1993J, made from 
year 1993 through the end of year 2005. 
There is a total of 
69 observing epochs, many of them made at several frequencies.
All these observations used global VLBI arrays. In nearly all 
of them (except for some epochs in 1993 and 1994), the whole VLBA 
(10 identical antennas of 25\,m diameter spread 
over the USA) and the Phased-VLA (equivalent to a $\sim$130\,m antenna in
New Mexico, USA) were used. Other antennas observed less often 
(each antenna participated in around 50\% of the epochs): 
Green Bank (110$\times$100\,m, West Virginia) and Goldstone (70\,m, California) in 
the USA, and part of the European VLBI Network (Effelsberg, 100\,m, Germany; 
Medicina, 32\,m, Italy; Noto, 32\,m, Italy; Jodrell Bank, 
76\,m, UK; Onsala, 25\,m, Sweden; Westerbork, 93\,m, The Netherlands; and Robledo, 70\,m, 
Spain). The arrays typically consisted of about 15 antennas at each observing epoch, 
with the exception of the first 4 epochs of Marcaide's group (between 1993 
and 1994, see Table 1) with less than 6 antennas.

\addtocounter{table}{1}   

At each epoch, the observations typically lasted between 12 and 16\,hr ( again, with 
the exception of some shorter runs of Marcaide's group in 1993 and 1994). 
The observations were taken in a phase-reference manner, with the exception of the
first ten epochs of Marcaide's group (the supernova was well detected in all baselines 
at these epochs). Scans of the radio core of source M\,81 
(hereafter, M\,81*) were inserted between the scans of the supernova, with duty cycles 
a few minutes long (from $\sim$2 to $\sim$10\,min, depending on epoch and 
observing frequency). Fringe finders and flux calibrators, both primary and 
secondary, were usually observed a few times during each epoch (depending on epoch, 
the sources \object{3C\,286}, \object{3C\,48}, \object{B0917+624}, \object{B0954+658}, 
and \object{OQ\,208} were observed).

Additional technical details on these observations can be found in    
Marcaide et al. (\cite{Marcaide1997}, \cite{Marcaide2009}), Bartel et al. 
(\cite{Bartel2002}), Mart\'i-Vidal (\cite{MartiVidal2008}), and references therein.

\subsection{Data calibration and imaging}
\label{Calibracion}

We completely re-calibrated (in amplitude, delay, and delay rate) all the available 
epochs of Bartel's group that were observed after 1995, and all our own 
phase-referenced epochs, following a uniform strategy. For the epochs earlier 
than 1995 that were observed by Bartel's group, we adapted the results published 
in Bartel et al. (\cite{Bartel2002}) to our analysis 
strategy, as we explain in Sect. \ref{IV}. For the visibility calibration, 
we used the NRAO Astronomical Image Processing System ({\sc aips})\footnote{
{\tt http://www.aips.nrao.edu}}. We first aligned 
the visibility phases through all the frequency
bands (for all sources and times) by fringe-fitting the single-band delays 
of one scan of a fringe finder, or flux calibrator, and then applying the estimated
antenna delays and phases to all visibilities. Then, a second fringe-fitting, now using 
the multiband delays, provided the new phase, delay, and rate corrections to all 
the observations. We performed the visibility amplitude calibration using system 
temperatures and gain curves from each antenna. We then transferred the calibrated 
visibilities of M\,81* into the program {\sc difmap}\footnote{
{\tt ftp://ftp.astro.caltech.edu/pub/difmap}} (Shepherd, Pearson \& Taylor 
\cite{Shepherd1995}) and 
made several iterations of phase and gain self-calibration until we obtained a 
high-quality image of M\,81*. The M\,81* CLEAN model obtained in {\sc difmap} was used 
again in {\sc aips} for another fringe-fitting iteration of the M\,81* data. Therefore, 
the new estimated antenna phases, delays, and rates were free of the (small) structure 
contributions of M\,81*. Such antenna corrections were then interpolated in time 
and applied to the SN\,1993J visibilities using the {\sc AMBG} option of the {\sc aips} 
task CLCAL\footnote{Using this option, {\sc aips} tries to find out the possible 2$\pi$ 
phase cycles introduced in the residuals between the scans of the calibrator source, 
correcting the phase interpolations of the target source scans.}. The amplitudes 
of the antenna gains were refined using the CLEAN model of M\,81* with the {\sc aips} 
task CALIB. These corrections were also interpolated and applied to the scans of 
SN\,1993J. For the case of Bartel's group observations, we did not apply any calibration 
to the cross-polarization data, and all our images and fits were performed using the 
Stokes I data (i.e., total intensity of the source, RCP + LCP flux densities).
At this stage, we edited bad visibilities based on standard selection criteria.

The process of imaging the supernova was performed following the special procedure 
described in Marcaide et al. (\cite{Marcaide2009}), but re-centering the supernova shell 
(at each frequency and epoch) according to the shifts reported in the next 
section. 
The most important details related to the imaging procedure described in Marcaide et 
al. (\cite{Marcaide2009}) are 1) use of a dynamic Gaussian taper in Fourier space 
prior to the deconvolution (to avoid possible resolution artifacts) and 2) 
phase self-calibration restricted to the shortest baselines, taking advantage of 
the source azimuthal symmetry. To check for any possible effect coming from this 
calibration procedure, we repeated all the analyses described in the following sections
using the phase-referenced visibilities directly, without any further calibration.  

We notice that the visibilities of epochs observed at 2.3\,GHz are noisier
than those of the other epochs. The {\em uv}-coverages of most of these epochs are also 
poor. Therefore, the quality of the results obtained at 2.3\,GHz is worse (see the 
dynamic ranges of all epochs in Table 1). In any case, the results at 2.3\,GHz are 
consistent with those obtained at the other frequencies.

\section{Data analysis}

\subsection{Location and proper motion of the expansion center}
\label{III}

The first step in the analysis process was to precisely determine the location 
of the supernova geometric center (i.e., the center of the shell-like structure) and
its possible evolution in time. This step is essential for a correct determination of the
shell size (and width), since a systematic offset of the fitted models from the real shell
center would translate into biases in the analysis strategies reported in this paper. 
To locate the supernova center, we fitted a simplified shell model to the 
visibilities. Such a model consisted of a homogeneous, optically-thin, spherical shell 
of outer radius $R_{\mathrm{out}}$ and inner radius $R_{\mathrm{in}}$ with a variable 
degree of central absorption by the ejecta (i.e., with partial blockage of the emission from 
the rear side of the shell for radii $< R_{\mathrm{in}}$).

Previous analyses of SN\,1993J data (Marcaide et al. \cite{Marcaide2005b}, Bietenholz et
 al. \cite{Bietenholz2003}), each based on a different approach, have concluded that 
there must be some opacity to the radio 
emission by the ejecta material. Marcaide et al. (\cite{Marcaide2005b}) conclude, 
from their Green-function deconvolution 
approach, that the ejecta opacity should be 100\% at all wavelengths, at least for 
the epochs they analyzed. Marcaide et al. (\cite{Marcaide2009}) confirmed these results by studying
the shapes of the azimuthal average of the supernova at different epochs and comparing 
them with several theoretical models. 
In contrast, Bietenholz et al. (\cite{Bietenholz2003}) conclude that the ejecta absorption
must be as small as 25\%. These authors used a simplified disk-like absorption 
pattern for the radio emission, a model we consider unrealistic. To 
accurately determine the location of the shell center, and to check for any possible 
biases coming from the use of different absorption models to fit the visibilities, we used 
a shell model with a variable degree of absorption from the ejecta and studied the 
effect of using different ejecta absorptions on the estimates of the shell center. 

\begin{figure}[t!]
\centering
\includegraphics[width=9cm]{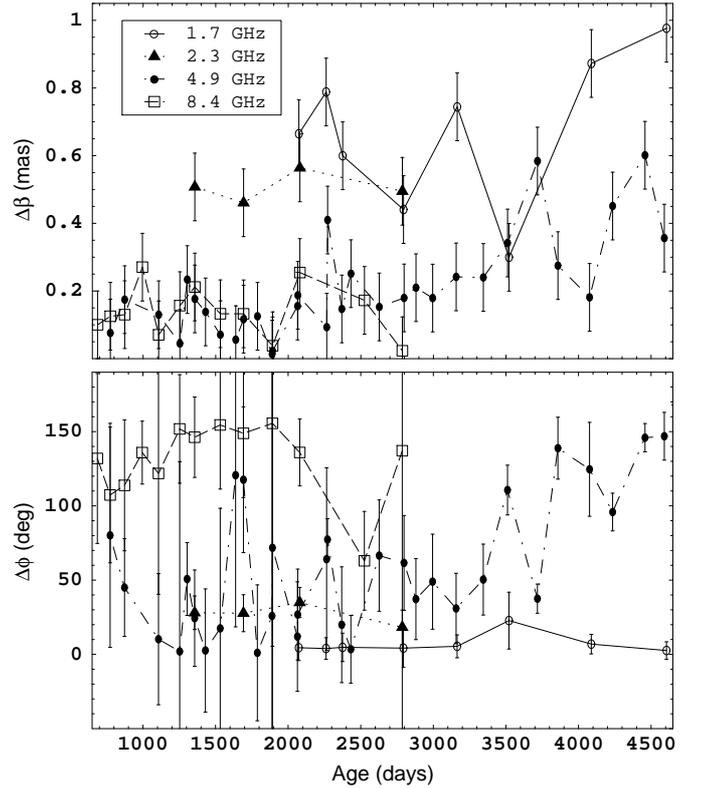}
\caption{Shifts in separation, $\beta$, and position angle, $\phi$, of the SN\,1993J 
shell center (using M\,81* as phase calibrator) with respect to the SN\,1993J position 
on day 1889 after explosion.}
\label{ExplosionCenter}
\end{figure}

The Fourier transform of a shell model with absorption from the inner side does
not have an analytical expression. Thus, we generated an 
interpolation function of such a Fourier transform, using the fractional shell width,
$ \xi = (R_{out} - R_{in})/R_{out}$, the shell radius, $R_{out}$, the percentage of 
absorption, and the position of the shell center as interpolating variables for 
computating the $\chi^2$ of our fits. This model is essentially the same as the one 
described in Appendix 2 of Marcaide et al. (\cite{Marcaide2009}), but with the 
addition of a plane-wave phase factor to account for the shift of the 
shell in the sky plane. Fitting the 
visibilities to a shell model with maximum ejecta absorption and a relative shell width of 
0.3, we estimated the shell size and the position of the supernova shell center for all 
epochs observed since year 1995. We notice that changing the shell width by a 20\% or 
removing the absorption by the ejecta from the model did not change the estimates of the 
coordinates of the shell center above the 1\,$\sigma$ level. The shifts obtained, 
taking the supernova position at 5\,GHz on day 1889 after explosion as reference, are shown 
in Fig. \ref{ExplosionCenter}.

From this figure, we reach several conclusions. The first is 
that the location of the phase-calibrator source, M\,81*, taking the shell center 
of SN\,1993J as a position reference, depends on the observing frequency.
In fact, we find an average shift of 0.62\,$\pm$\,0.04\,mas in $\alpha\,\cos{\delta}$ 
and 0.10\,$\pm$\,0.02\,mas in $\delta$ between the M\,81* positions at 5\,GHz 
and 1.7\,GHz, the average shift between 5\,GHz and 2.3\,GHz is 0.40\,$\pm$\,0.04\,mas 
in $\alpha\,\cos{\delta}$ and 0.2\,$\pm$\,0.1\,mas in $\delta$, and the shifts between 5\,GHz 
and 8.4\,GHz are smaller ($-$0.13\,$\pm$\,0.3\,mas in $\alpha\,\cos{\delta}$ and 
$-$0.01\,$\pm$\,0.03\,mas in $\delta$). 

According to the 
standard jet model (Blandford \& K\"onigl \cite{Blandford1979}), there should indeed be
spectral shifts in 
the brightness peaks of the VLBI core-jet structures due to the frequency-dependent 
transition of the ejected material from optically thick to optically thin, for the 
synchrotron radiation, as first found by Marcaide \& Shapiro (\cite{Marcaide1983}, 
\cite{Marcaide1984}) and later confirmed in many cases (e.g. Kovalev et al. 
\cite{Kovalev2008}). Similar results of the spectral shift of M\,81* were reported by 
Bietenholz et al. (\cite{Bietenholz2004}), 
who have estimated the position of the real core (i.e., the central black hole) of M\,81* by 
finding a sharp bound in the radio emission at all frequencies, taking these shifts and 
the different sizes of the radio structures into account. Our in-depth analysis of the 
M\,81* structure evolution and core location at several frequencies and epochs 
will be reported elsewhere.

Another conclusion extracted from Fig. \ref{ExplosionCenter} is that the location
of the supernova shell center does not evolve in time with respect to 
the phase calibrator, at least at our level of precision. There is a hint of a drift 
at 5\,GHz beyond day $\sim$3000, but the scatter of data at these epochs 
is also larger. The proper motion at each frequency, fitted from the position shifts shown 
in Fig. \ref{ExplosionCenter}, 
is compatible with zero at a 1-sigma level. The most siginificant value of proper motion 
is obtained if we fit the 5\,GHz shifts from day 3000 onwards. In this case we obtain a 
proper motion with module $54\pm31$\,$\mu$as\,yr$^{-1}$, which is compatible with zero 
at a 2-sigma level. Bietenholz et al. (\cite{Bietenholz2001}) also 
studied the proper motion of the SN\,1993J radio shell and arrived at the conclusion that it 
is compatible with zero at the level of precision achievable with VLBI. To take 
the shift in the peak of M\,81* into account at different frequencies, we corrected the 
visibilities of SN\,1993J at each epoch by applying the average shifts found in the 
SN\,1993J images between frequencies, prior to any further data analysis.
The center of the shell is thus a fixed parameter in all the analyses reported in the 
following sections\footnote{{For the observations that were not phase-referenced to 
M\,81*, we used the shell center estimated from model fitting.}}. 

We notice that the scattering in the estimated positions of SN\,1993J, shown
in Fig. \ref{ExplosionCenter}, may translate into an additional uncertainty in the estimate
of the size of the supernova shell, since we consider the position of M\,81* stationary 
with respect to the supernova shell center at each frequency. In the worse cases (i.e., 
smallest shell sizes), the standard deviation in the SN\,1993J coordinates is around 
$10-20$\% of the supernova radius. Such a large shift in the supernova shell center could 
be easily appreciated by visual inspection in the images. We did not see such large deviations 
in the position of the supernova shell center at any of the epochs reported here. Additionally, 
even if such large shifts had taken place in the estimate of the supernova shell 
center, as determined from phase-referencing to M\,81*, we conclude from simulations that 
the effect of such (random-like) shifts on the shell size would be on the order of 5\%, 
which translates into an effect of only 1\%, or lower, on the fitted parameters of the 
expansion curve reported in the next section.

\subsection{Expansion curve}
\label{IV}

In this section, we report on the combined analysis of the available VLBI data with different 
approaches. On the one hand, we applied a novel analysis method to the data in the sky plane 
(the common-point method, CPM, described in Mart\'i-Vidal \cite{MartiVidal2008} and Appendix A of 
Marcaide et al. \cite{Marcaide2009}). This method relies on some mathematical properties of 
the convolution of a Gaussian with an image of a source with azimuthal symmetry. It can be shown 
that there are points in the azimuthal average of the convolved image that do not change under 
first-order modifications of the width of the convolving Gaussian. These points can be related 
to the size of the source. 

On the other hand, we used model fitting to the 
visibilities, following the same approach as in Marcaide et al. (\cite{Marcaide2009}) (see 
their Sect. 4.4 and their Appendix B). This approach for model fitting has some advantages, compared
to the approach followed by Bartel et al. (\cite{Bartel2002}) (see Sect. 8 of Marcaide et al. 
\cite{Marcaide2009} for more details). Then, we check the similarities and 
discrepancies of the expansion curves obtained with both approaches.

\subsubsection{Expansion with the common-point method}
\label{ExpansionCPM}

The CPM analysis of all the images of SN\,1993J obtained as described in Sect. \ref{II}, 
yields the shell radii shown in Table 1, 
Col. 5 (we call these size estimates $R_{\mathrm{SC}}$). For epochs earlier than 20 
September 1994, the supernova sizes cannot be measured well with the CPM. It would imply 
over-resolution of the images. Indeed, the convolving beam after applying the CPM to 
the data of epoch 23 February 1995 is 0.74\,mas, which is similar to, but still slightly smaller 
than, the corresponding interferometric beam ($\sim$0.94\,mas using uniform weighting).
Therefore, for data observed before 1995, we used the shell sizes estimated from 
model fitting to the visibilities. For epochs of Bartel's group earlier than 
year 1995, we used their results published in 
Bartel et al. \cite{Bartel2002}. These sizes were then transformed into ``CPM sizes'' by applying 
the corresponding factors, which were obtained theoretically from numerical simulations.
For a shell with a fractional width of 0.3 and maximum absorption from the ejecta 
(i.e., the emission structure that we assume for SN\,1993J), the ratio of model-fitting to 
CPM size\footnote{This factor is computed for a shell with radius of 0.6\,mas, 
observed at the frequency of 5\,GHz. This factor slightly depends on the shell size and/or on
the {\em uv}-coverage.} is 1.031, provided the 
model used to fit the visibilities does not take ejecta absorption into account and has a 
fractional width of only 0.2 (i.e., the model used in Bartel et al. \cite{Bartel2002}).
In any case, all these 
transformation factors are always close to unity. Indeed, their deviations from 1 are similar to 
the relative uncertainties of the shell-size estimates. 

Also, for very early epochs, when the synchrotron self-absorption 
is large, a shell profile is not able to properly model the supernova structure, since the 
emission pattern of a spherically symmetric optically thick source is disk-like. Moreover, the 
transition from optically thick to optically thin for the synchrotron radiation, 
is frequency-dependent. Therefore, we expect to have some frequency-dependent 
biases in the expansion curve determined for very early epochs (earlier than day $\sim 100-200$ 
after explosion). 
However, these few data do not affect our results above the 0.5\,$\sigma$ level, given the 
long time coverage and dense sampling of the expansion curve.

\begin{figure*}[ht!]
\centering
\includegraphics[width=12cm,angle=0]{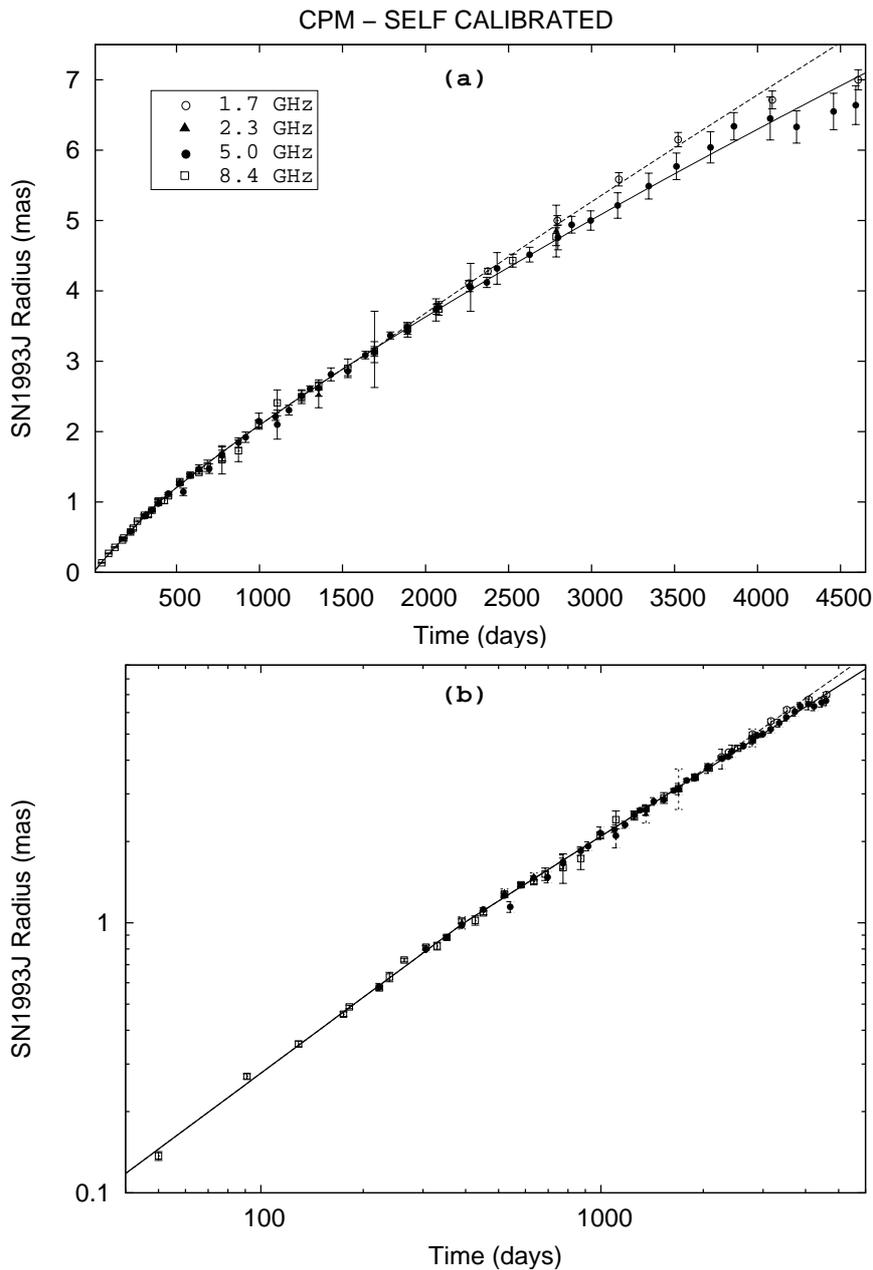}
\caption{(a) Expansion of supernova SN\,1993J. The radii were measured with the CPM 
applied to the images obtained from the process described in Sect. \ref{II} 
(we call these sizes $R_{\mathrm{SC}}$). The 
dashed line corresponds to the model given by Eq. \ref{ExpanNoBreak}, fitted to only 
the 1.7\,GHz data. The continuous line is the model given by Eq. \ref{ExpanBreak}, 
fitted to the data at higher frequencies. (b) Same as in (a), but in logarithmic scale.}
\label{ExpanFigure}
\end{figure*}

Figure \ref{ExpanFigure}(a) (plot on a linear scale) and (b) (plot on a 
logarithmic scale) shows several weighted-least-square fits of the supernova 
radius, $R$, to power laws of time. The last four epochs were excluded from the fits, 
since they clearly depart from the general behavior. The fit shown as a continuous line 
uses all data at 22, 15, 8.4, and 5\,GHz, and fits a model with 4 parameters: two 
expansion indices ($m_1$ and $m_2$), a break time ($t_{\mathrm{br}}$, that separates the 
expansion regimes given by each expansion index), and the supernova size at the break time 
($r_{\mathrm{br}}$). The expression for the fitted model is

\begin{equation}
R(t) = \left\{\begin{array}{rl} 
r_{\mathrm{br}}\,\left(t/t_{\mathrm{br}}\right)^{m_1},~~~ t < t_{\mathrm{br}}  \\
r_{\mathrm{br}}\,\left(t/t_{\mathrm{br}}\right)^{m_2},~~~ t \geq t_{\mathrm{br}}.
\end{array} \right. 
\label{ExpanBreak}
\end{equation}

This model is the same as the one used by Marcaide et al. (\cite{Marcaide2009}) to fit
their data at 8.4 and 5\,GHz. The uncertainties of the measured radii were uniformly scaled 
to make the reduced $\chi^2$ equal to unity (the resulting uncertainties are those shown in 
Table 1 
Col. 5). The results of the fit to data at 22, 15, 8.4, and 5\,GHz are
shown in  Table \ref{SNExpanAll}, row 1. 
Adding the data at 2.3\,GHz to the fit does not change the results above the 1\,$\sigma$ 
level. 

Since we adapted the published (model fitting) size estimates to CPM size estimates 
for epochs earlier than day 541, one could suspect that the fitted break time (and, 
consequently, also $m_1$ and $m_2$) could be strongly affected by the conversion factors applied
to the model-fitting sizes, in order to {\em convert} them into CPM sizes. This is not the case;
indeed, since using very different emission models (shell widths ranging from 0.2 to 0.3 and/or 
different levels of ejecta opacity, from 100\% to 0\%) translates into deviations in the 
modelfit-to-CPM conversion factors of $\sim3$\%, we multiplied all the sizes of epochs earlier 
than day 541 first by 0.97, and later by 1.03, and each time re-fitted the resulting expansion 
curves. The resulting expansion parameters are compatible each time to those shown in Table 
\ref{SNExpanAll}, row 1, at a $\sim0.4\sigma$ level (for $t_{br}$), $\sim0.2\sigma$ level 
(for $m_1$), and at a $\sim1.2\sigma$ level (for $m_2$). The expansion parameters shown in 
Table \ref{SNExpanAll} are, therefore, very insensitive to the conversion factors applied to 
the (early) model-fitting sizes.

The fit shown in Fig. \ref{ExpanFigure} (dashed line) uses data only at 1.7\,GHz, 
and the simple model given by equation 

\begin{equation}
R(t) \propto t^{m_3}.
\label{ExpanNoBreak}
\end{equation}

\noindent The uncertainties of the measured radii were also scaled to make the reduced 
$\chi^2$ equal to unity. The result of the fit to only the 1.7\,GHz data is
shown in Table \ref{SNExpanAll}, row 1. 

To quantify the evidence of the frequency-dependent expansion described in 
Marcaide et al. (\cite{Marcaide2009}) and in this paper, we used the model given 
by Eq. \ref{ExpanNoBreak} to fit the size estimates at 5\,GHz for epochs later than 
day 2000 (i.e., roughly the supernova age at the first 1.7\,GHz epoch).
The resulting expansion index at 5\,GHz is $m = 0.771\pm0.016$, which is at 
$\sim2.8\sigma$ from the expansion index determined at 1.7\,GHz in the same time 
range ($m=0.87\pm0.02$).

\begin{figure*}[ht!]
\centering
\includegraphics[width=12cm,angle=0]{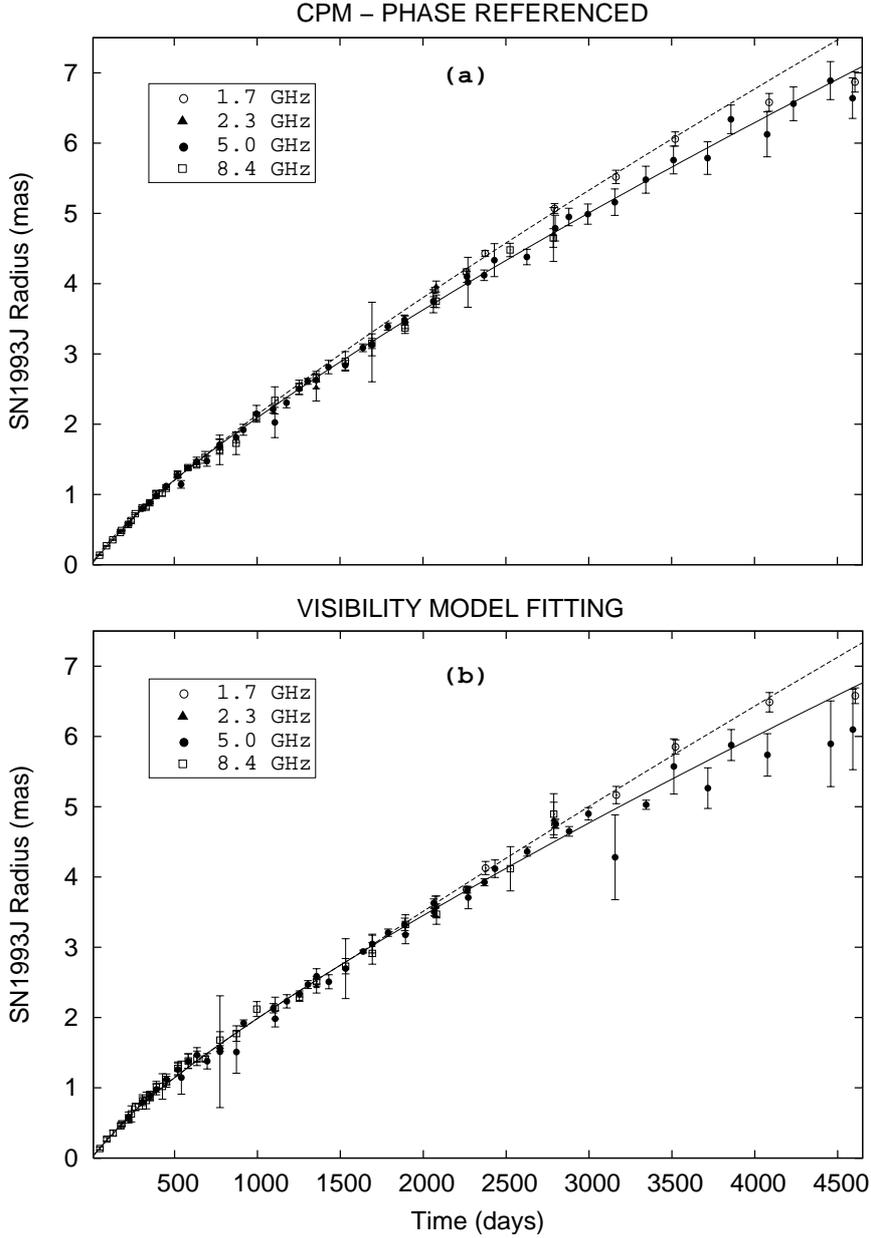}
\caption{(a) Same as in Fig. \ref{ExpanFigure}(a), but with radii measured 
with the CPM applied to the images obtained from the phase-referenced visibilities 
(we call these size estimates $R_{\mathrm{PR}}$). (b) Same as in Fig. 
\ref{ExpanFigure}(a), but with radii estimated from model fitting to the 
visibilities (we call these estimates $R_{\mathrm{MF}}$).}
\label{ExpanFigUVFIT}
\end{figure*}

We also applied the CPM to the supernova images obtained just after the phase-reference
calibration (i.e., without any phase self-calibration). The resulting expansion curve
and fitted parameters are shown in Fig. \ref{ExpanFigUVFIT}(a) and Table \ref{SNExpanAll} 
(row 2), respectively. We notice that these results are compatible with those given just 
above.

We should also point at that even though these fits describe the supernova expansion 
satisfactorily for most of the 
observations, for epochs observed after day $\sim3500-4000$ the sizes at all frequencies are 
systematically smaller than the model predictions, as can be seen in Figs.
\ref{ExpanFigure} and \ref{ExpanFigUVFIT}. Even then, the observed size at 1.7\,GHz on 
day 4606 after explosion is larger than the size at 5\,GHz observed on day 4591. In 
Paper II, we will discuss this ``enhanced'' deceleration of the supernova shell at very late 
epochs and its relationship with the exponential drop of the radio light curves reported by 
Weiler et al. (\cite{Weiler2007}).

\subsubsection{Expansion from analysis in Fourier space}
\label{ExpanFourier}

If we fit a partially-absorbed shell model (the model used in Sect. \ref{III} and described in 
Marcaide et al. \cite{Marcaide2009}, Appendix B) 
to the 22, 15, 8.4, and 5\,GHz visibilities obtained after the calibration described in 
Sect. \ref{II} (we call these size estimates $R_{\mathrm{MF}}$), 
the resulting expansion curve is the one shown in Fig. \ref{ExpanFigUVFIT} (data of day 4235 at 
5\,GHz could not be properly modeled). The fitted parameters 
are shown in Row 3 of Table \ref{SNExpanAll}. 
Adding the data at 2.3\,GHz to the fit does not change these results above the 
1\,$\sigma$ level. The result of the fit to only the 1.7\,GHz data is also shown in 
Row 3 of Table \ref{SNExpanAll}. 

\begin{figure}[t!]
\centering
\includegraphics[width=9cm,angle=0]{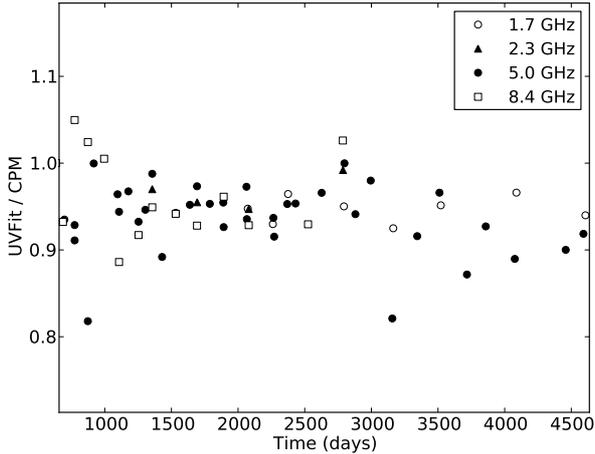}
\caption{$R_{\mathrm{MF}}/R_{\mathrm{SC}}$, i.e., ratios of sizes estimated from visibility 
model fitting ($R_{\mathrm{MF}}$, see Sect. \ref{ExpanFourier}) to sizes obtained with the CPM 
($R_{\mathrm{SC}}$, see Sect. \ref{ExpansionCPM}).}

\label{SizeRatios}
\end{figure}

The parameters fitted to the expansion curve determined with the CPM are compatible 
with those of the expansion curve obtained from visibility model fitting, although the 
scatter in the latter expansion curve is higher. In Fig. \ref{SizeRatios}, 
we show the ratios of sizes estimated
from visibility model fitting to those estimated with the CPM. These ratios are $\sim0.95$, and 
the scatter in this plot is mainly due to the scatter in the sizes estimated from visibility 
model fitting. The ratios at 1.7\,GHz are less scattered, and the ratios at 5\,GHz seem to 
be slightly lower at later epochs. This is an expected result if the different sizes at 
different frequencies are due to different radial intensity profiles of the supernova
(see Sect. \ref{Spix}). Since the effect of different intensity profiles is lower for the CPM 
sizes than for the model-fitting sizes (i.e., the CPM size estimates are more insensitive to 
changes in the radial intensity profile of the source, see Appendix A of Marcaide et al. 
\cite{Marcaide2009}), any systematic effect in the data towards a smaller size estimate 
(like a decrease in the ejecta opacity) should decrease the ratio of model-fitting sizes 
to CPM sizes.

\subsection{Comparison of the different expansion models}
\label{ComparSec}

\begin{table*}[th!]
\begin{minipage}[t]{17cm}
\caption{Expansion parameters of SN\,1993J.} 
\label{SNExpanAll}
\centering
\renewcommand{\footnoterule}{}
\begin{tabular}{  l | r@{ $\pm$ }l   r@{ $\pm$ }l   r@{ $\pm$ }l  | r@{ $\pm$ }l } 
\hline\hline
 & \multicolumn{6}{c}{ 5 to 22\,GHz} & \multicolumn{2}{|c}{1.7\,GHz}\\
 & \multicolumn{2}{|c}{$m_1$} & \multicolumn{2}{c}{$m_2$} & 
\multicolumn{2}{c}{$t_{br}$ (days)} & \multicolumn{2}{|c}{$m_3$}\\
\hline
CPM selfcal\footnote{Using CPM-measured supernova sizes from images obtained 
from self-calibrated visibilities, as described in Marcaide et al. (\cite{Marcaide2009}).}
& 0.933 & 0.010 & 0.796 & 0.005 & 390  & 30  & 0.87 & 0.02 \\
CPM ph-ref\footnote{Using CPM-measured sizes from images obtained from phase-referenced 
visibilities.}     
& 0.933 & 0.010 & 0.795 & 0.005 & 390  &  40  & 0.83 & 0.02 \\
Model fitting\footnote{Using model fitting to the visibilities.}  
& 0.94  & 0.06  & 0.798 & 0.007 & 270  & 70  & 0.90 & 0.03 \\ 
Bartel 1\footnote{Using the shell sizes reported in Bartel et al. (\cite{Bartel2002}).} 
& 0.93  & 0.02  & 0.798 & 0.006 & 390  & 50  & 0.84 & 0.06 \\
Bartel 2\footnote{Using the sizes reported in Bartel et al. (\cite{Bartel2002}), but taking 
only the epochs later than day 182 after explosion (i.e., the day of the first epoch reported 
in Marcaide et al. \cite{Marcaide2009}).} 
& 0.82  & 0.03  & 0.796 & 0.016 & 1000 & 700 & 0.84 & 0.06 \\
Marcaide\footnote{Refit using the shell sizes reported in Marcaide et al. (\cite{Marcaide2009}).}
& 0.845 & 0.005 & 0.788 & 0.015 & 1500 & 300 & 0.87 & 0.03 \\

\hline
\end{tabular}
\end{minipage}
\end{table*}

In Table \ref{SNExpanAll} we summarize the results of fitting the models given by 
Eqs. \ref{ExpanBreak} and \ref{ExpanNoBreak} to the expansion curves of SN\,1993J,
obtained with different approaches and different time coverages.
The results shown in the first three rows of Table \ref{SNExpanAll} correspond to 
fits using the complete set of VLBI data here reported, analyzed following 
the approaches described in the previous sections. The first row corresponds 
to the sizes determined with the CPM, applied to the supernova images obtained following the 
special self-calibration described in Marcaide et al. (\cite{Marcaide2009}). The analysis 
procedure described in that publication has passed several tests with synthetic data and has 
been shown to give more precise size estimates than model fitting to the visibilities. 
Therefore, the expansion model that we consider definitive for SN\,1993J is the one 
corresponding to the first row of Table \ref{SNExpanAll}.

The fitted parameters in the fourth row correspond to the supernova sizes published in 
Bartel et al. (\cite{Bartel2002}). These parameters are very close
to the results reported here for the complete set of observations (rows 1, 2, and 3). 
Therefore, even though the conclusions of Bartel et al. (\cite{Bartel2002}) are very different 
from ours, their fitted shell sizes are compatible with our expansion model, i.~e., one
expansion regime for the 1.7\,GHz data and two expansion regimes (separated by a break
time) for the higher frequencies. Indeed, the sizes at 1.7\,GHz reported in 
Bartel et al. (\cite{Bartel2002}) are up to 5\% larger than those reported by the same 
authors at the higher frequencies. In our analysis, the maximum size difference between 
1.7 and 5\,GHz is $\sim7.3$\% (observations of November 2001).

The last two rows of Table \ref{SNExpanAll} correspond to fits of the subset of 
observations reported in Marcaide et al. (\cite{Marcaide2009}) and the subset 
of observations reported by Bartel et al. (\cite{Bartel2002}), but taking only 
those from day 182 after explosion onwards (for a direct comparison of
the models fitted to the data reported in these two works). 
We note that Marcaide et al. (\cite{Marcaide2009}) claim that the sizes at 
1.7\,GHz can be modeled by an extrapolation of the early expansion curve 
(i.e., that with the expansion index $m_1 = 0.845$). Thus, a fit to the 1.7\,GHz 
sizes, alone, was not reported in Marcaide et al. (\cite{Marcaide2009}). Fitting 
those 1.7\,GHz sizes using Eq. \ref{ExpanNoBreak} results in $m_3 = 0.87$ (Table \ref{SNExpanAll}, 
row 6), which is within 1$\sigma$ from $m_1$.

As can be seen, the fits shown in the first four rows are very similar. The 
results in the last two rows differ from the other ones mainly in 
the estimates of the break time (which also has large statistical uncertainties) and 
of $m_1$ (which is $\sim$9\% lower). Clearly, the poor early time coverage of the 
subsets of observations corresponding to the fits of rows 5 and 6 of Table \ref{SNExpanAll} 
results in a lower $m_1$, which in turn results in a later $t_{\mathrm{br}}$ to 
adequately model the later supernova expansion. 

We could try to fit our complete expansion curve 
to a model with two break times, to check whether the break of day 1500 reported by Marcaide et al.
(\cite{Marcaide2009}) (and reproduced here in row 6 of Table \ref{SNExpanAll}) can or cannot be 
recovered from the analysis of the whole data set. However, 
we find from Monte Carlo simulations that fitting such a model with two breaks to the data 
would not give reliable results (as long as such breaks are left as free parameters in the fit). 
Therefore, it is difficult to conclude whether both 
breaks are present in the data or not. We notice, however, that Marcaide et al. 
(\cite{Marcaide2009}) interpret the break time reported at day 1500, not as a real break in the 
expansion curve, but as the result of several effects (related to the ejecta opacity and/or to a 
possible radial drop in the amplified magnetic fields inside the emitting region), which are 
strong enough to affect the measured expansion curve at the higher frequencies, but not at 
1.7\,GHz. This interpretation of the expansion curve, taking also the early break at day $\sim400$ 
into consideration, will be analyzed in Paper II.

\subsection{Structure evolution of SN\,1993J}
\label{V}

\subsubsection{Estimates of ejecta opacity and shell width}

We measured the supernova shell radius using a method unrelated to 
model fitting in Fourier space (i.~e., the CPM). Thus, we can use these 
measured sizes as fixed parameters in a model fitting, in which we can estimate the 
fractional shell width. Given that the CPM has a small bias, dependent on the emission 
structure of the supernova (see Marcaide et al. \cite{Marcaide2009}), we 
should correct the shell sizes with the right bias before estimating the shell 
width. However, the bias of the CPM depends on the degree 
of absorption by the ejecta, and also on the fractional shell width, which is the quantity 
to be determined from model fitting. In short, we have a coupling between fitted shell 
widths and CPM biases. We can 
look for self-consistency in that coupling, finding a shell width for which the bias of 
the CPM, applied to the (fixed) shell sizes in the model fitting, translates into a fitted 
shell width corresponding to the CPM bias already applied. See Sects. 6.2 and 7.2 of 
Marcaide et al. \cite{Marcaide2009} for a detailed description 
of the trial-error procedure to find this self-consistency. Such a self-consistent 
fractional shell width is $\sim$0.35, for a model with maximum (i.e., 100\%) ejecta opacity. 

The percentage of ejecta opacity could be somewhat lower than 100\% 
and/or could evolve in time. When we require self-consistency between 
(bias-corrected) CPM results and (model-fitted) shell widths, lower ejecta opacities 
translate into narrower shell widths. In that sense, a value of 0.35 can be 
considered as an observational upper bound of the fractional shell width of SN\,1993J. 

An appropriate percentage of absorption to use in our fits is the average of the 
estimates obtained from fitting a shell model to the data of our best epochs (i.e., 
with good {\em uv}-coverages and large signal-to-noise ratios). The epochs selected for 
such fits were all at 5\,GHz between days 1638 and 2369 after explosion. We used 
data only at 5\,GHz to avoid any possible frequency-dependent bias. 
For all 10 epochs, we fitted the fractional shell width, 
$\xi = (R_{\mathrm{out}}-R_{\mathrm{in}})/R_{\mathrm{out}}$, the supernova radius, 
$R_{\mathrm{out}}$, the location 
of the shell center, the percentage of absorption, and the total flux density, obtaining 
an average relative shell width of 0.31\,$\pm$\,0.02 and an average percentage of 
absorption of (80\,$\pm$\,8)\%. These results are compatible with those reported in 
Marcaide et al. (\cite{Marcaide2009}), who used a subset of the observations presented here. 

\begin{figure}[t!]
\centering
\includegraphics[width=9cm,angle=0]{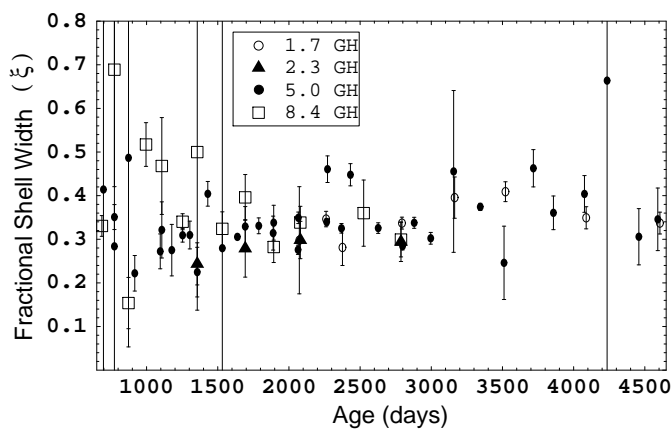}
\caption{Fitted fractional shell widths of SN\,1993J, using the model described 
in Appendix 2 of Marcaide et al. (\cite{Marcaide2009}) and the shell sizes 
determined with the CPM, applying the corresponding bias for an ejecta absorption 
of 80\% (see text). The errors shown are such that the reduced $\chi^2$
of each fit is equal to unity.}
\label{SHELLWIDTH}
\end{figure}

For an ejecta opacity of 80\%, the shell widths obtained from model 
fitting are shown in Fig. \ref{SHELLWIDTH}. Only observations of epochs from year 1995 
onwards were used in the fitting. Using these 
values, we obtain a weighted mean of the shell width of 0.310\,$\pm$\,0.011 for 8.4\,GHz 
data, 0.300\,$\pm$\,0.005 for 5\,GHz data, 0.26\,$\pm$\,0.02 for 2.3\,GHz data, and 
0.324\,$\pm$\,0.008 for 1.7\,GHz data.

All these quantities are close to 0.3. Based on very few and noisy data, the 
shell widths estimated at 2.3\,GHz are the smallest, within 2$\sigma$ 
from the value 0.3. On the other hand, data at 8.4 and 1.7\,GHz give wider 
shell width estimates than at 5\,GHz. The shell width at 8.4\,GHz is 
compatible with that at 5\,GHz at a 1\,$\sigma$ level. The average shell 
width at 1.7\,GHz is about 2$\sigma$ wider than at 5\,GHz. The difference
between shell widths at these frequencies could be due to either a physically 
wider shell at 1.7\,GHz, to instrumental effects related to the finite sensitivity
of the interferometers, or to a lower ejecta opacity at 5\,GHz (lower opacities 
translate into narrower fitted shell widths). Any of 
these explanations (or a combination of them) could help explain the different 
shell widths obtained for different frequencies (see Marcaide et al. 
\cite{Marcaide2009}). In Paper II, we will analyze these possibilities and their 
relationship with features in the radio light curves 
published by Weiler et al. (\cite{Weiler2007}).
From Fig. \ref{SHELLWIDTH}, we also notice that no time evolution of the relative 
shell width is discernible at any frequency, although the data are too noisy to 
reach any robust conclusion.

\begin{figure}
\centering
\includegraphics[width=10cm]{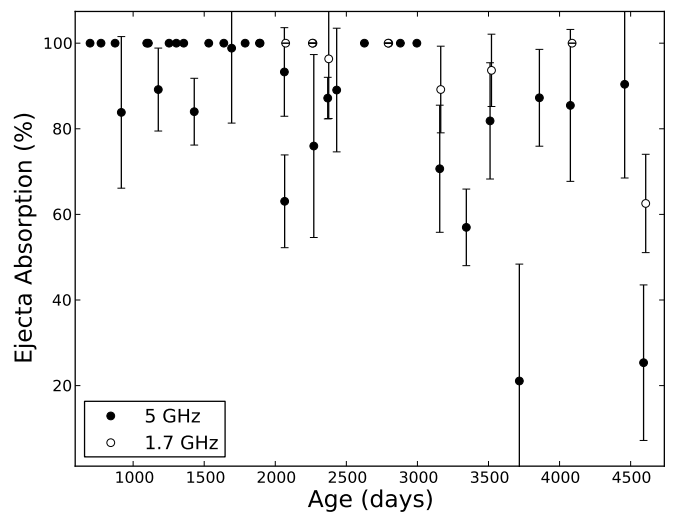}
\caption{Percentage of absorption (or degree of ejecta opacity) fitted to the 
visibilities by fixing the shell size and the fractional shell width (see text). The 
maximum allowed absorption (or ejecta opacity) in the model is set to 100\%.}
\label{AbsorptionFigure}
\end{figure}

For completeness, we fitted the percentage of absorption at 5\,GHz and 
1.7\,GHz by fixing the fractional shell width to 0.3 and the shell sizes to the 
estimates given by Eq. \ref{ExpanNoBreak} (according to our hypothesis of different 
ejecta absorptions at different frequencies, this equation should give the closest 
estimates of the true shell size at late epochs). The results obtained are shown 
in Fig. \ref{AbsorptionFigure}. We set the maximum possible absorption to 100\%, 
to obtain fits with physical meaning for some epochs. There is a hint of a larger 
absorption at 1.7\,GHz compared to that at 5.0\,GHz, although the data are too 
noisy to infer any clear evolution in the opacity.

\subsubsection{Spectral-index gradients in the shell}
\label{Spix}

Since the expansion curve of SN\,1993J is dependent on the observing frequency,
there must be a distribution of spectral indices through the shell. If the size
at 1.7\,GHz is really larger than at 5\,GHz, the spectral index in the outer part 
of the shell should tend to $-\infty$, since there would be emission at 1.7\,GHz but 
not at 5\,GHz. On the contrary, if the frequency effects in the expansion curve came from 
changes in the opacity by the ejecta, the spectral indices in the inner part 
of the shell would be larger than those in the outer shell (since the intensity at 5\,GHz 
in the inner shell would be higher, because of the lower ejecta opacity). 

\begin{figure}
\centering
\includegraphics[width=10cm]{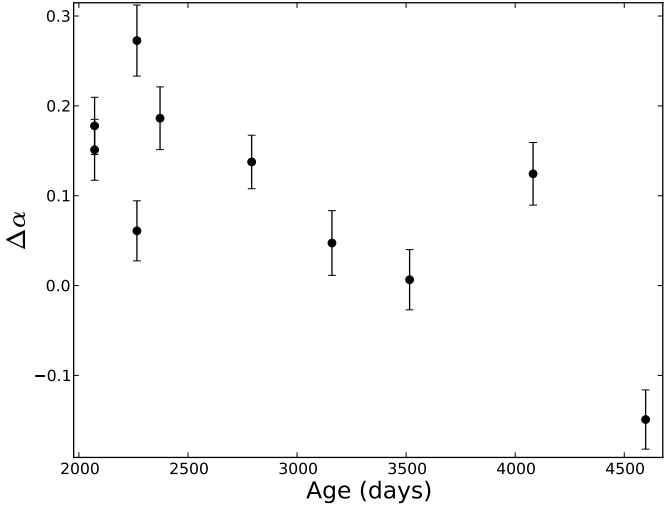}
\caption{Difference between the $1.7-5$\,GHz spectral index in the inner shell (i.e., for radii 
up to 0.5 times the shell radius) and in the outer shell (i.e., for radii larger than 0.5 
times the shell radius).}
\label{SpixEvol}
\end{figure}

Unfortunately, the spectral-index images of SN\,1993J between 1.7 and 5\,GHz are very noisy. 
No clear conclusion can be extracted from the images themselves. 
However, we can increase the SNR of the spectral-index estimates by integrating the intensity 
at each frequency through different parts of the shell. On the other hand, we integrate the flux 
densities from a radial distance of 0 (i.e., the shell center) up to half the shell size at 1.7\,GHz; 
we call these integrated flux densities $F^{in}_{5.0}$ and $F^{in}_{1.7}$, for the images at 5.0 
and 1.7\,GHz, respectively. On the other hand, we integrate the flux densities from a radial 
distance of 0.5 times the shell size at 1.7\,GHz up to 2 times the shell size at 1.7\,GHz (to be 
sure that we integrate all the emission from the outer shell at both frequencies); we call these 
integrated flux densities $F^{out}_{5.0}$ and $F^{out}_{1.7}$, for 5\,GHz and 1.7\,GHz, 
respectively. In Fig. \ref{SpixEvol} we show the difference between the inner spectral index, 

$$\alpha_{in} = \log{\left(\frac{F^{in}_{5.0}}{F^{in}_{1.7}}\right)}/\log{\left( \frac{5.0}{1.7} \right)}, $$

\noindent and the outer spectral index,

$$\alpha_{out} = \log{\left(\frac{F^{out}_{5.0}}{F^{out}_{1.7}}\right)}/\log{\left( \frac{5.0}{1.7} \right)}.$$

We call this difference $\Delta\alpha = \alpha_{in}-\alpha_{out}$. It can be seen in the figure 
that $\Delta\alpha$ is positive for all the epochs where quasi-simultaneous 1.7 and 5\,GHz 
observations were available (except for the last one, at day 4606 after explosion). This is
independent observational evidence of the opacity effects by the ejecta suggested in Marcaide et al.
(\cite{Marcaide2009}) and supported here. The values of $\Delta\alpha$ at the earlier epochs gather
around $0.1-0.3$. The expected value of $\Delta\alpha$ for a shell model with a fractional width
of 0.3 and zero absorption at 5\,GHz (and maximum at 1.7\,GHz) is 0.4, a value higher than those shown
in Fig. \ref{SpixEvol}. It should be noticed that the presence of hot spots in the outer shell and/or 
changes in the ejecta opacity through the inner shell at each frequency (and, of course, the different 
sampling of Fourier space between both frequencies) may affect the estimates
of $\Delta\alpha$.

\subsubsection{Azimuthal evolution of the shell inhomogeneities}

\begin{figure}[t!]
\centering
\includegraphics[width=9cm,angle=0]{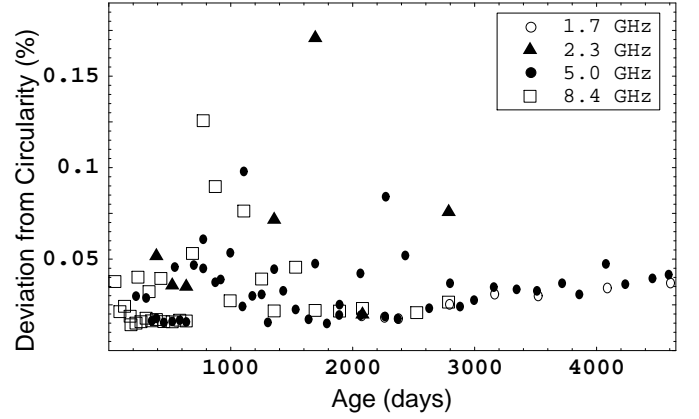}
\caption{Degree of circularity of the SN\,1993J shell, as indicated by the fractional 
uncertainty in the source radius estimates when determined with the CPM (see Appendix 1 of 
Marcaide et al. \cite{Marcaide2009}).}
\label{SN93J-circul}
\end{figure}

The VLBI images of SN\,1993J keep a high degree of circularity during 
the whole expansion. A quantitative representation of the degree of circularity of the shell 
is the fractional uncertainty of the radius determined with the CPM (i.~e., the scatter of 
radial positions of a given contour with respect to the shell center, in units of the source 
radius; see Appendix 1 of Marcaide et al. \cite{Marcaide2009} for more details). The degree
of circularity of the supernova, computed this way, is typically around 2-4\%, as can 
be seen in Fig. \ref{SN93J-circul}, with the exception of some epochs with low 
dynamic ranges. Similar results were also reported in Bietenholz et al. 
(\cite{Bietenholz2001}). Such a circularity in the images must be due to a high degree of 
isotropy in the angular distribution of the ejecta velocities (and, therefore, in the 
distribution of the CSM).

\begin{figure}[th!]
\centering
\includegraphics[width=8.5cm,angle=0]{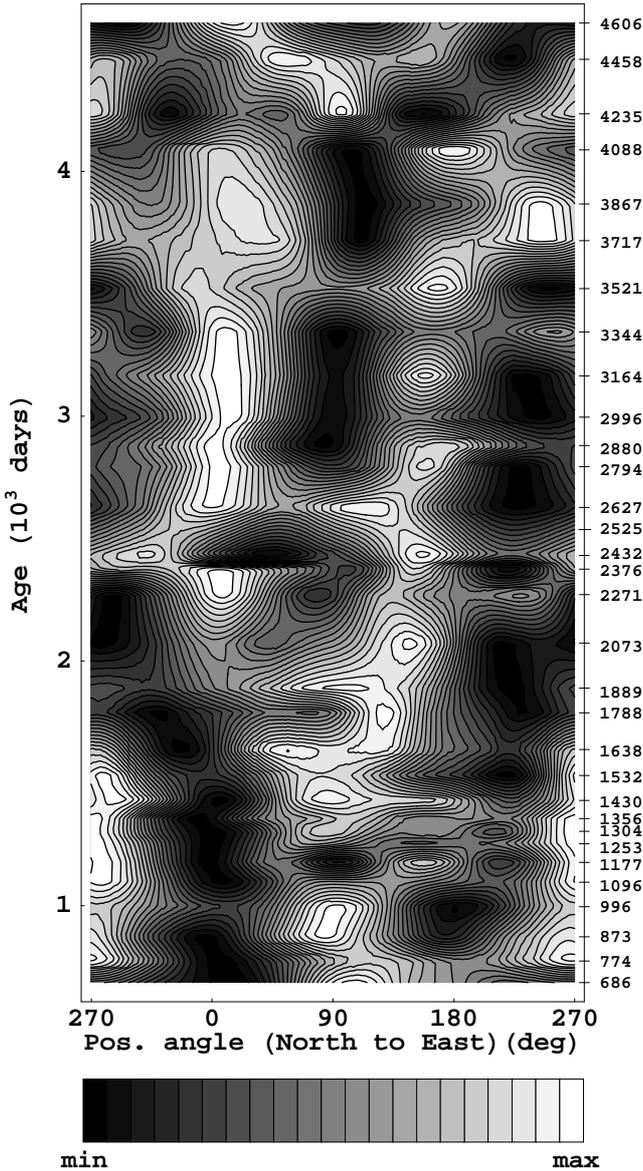}
\caption[Azimuthal evolution of SN\,1993J]
{Evolution of the azimuthal intensity distribution in the SN\,1993J shell, computed as 
the linear time interpolation of a selection of epochs (see text). The supernova age at 
the selected epochs is marked with ticks (right). The 20 contours shown are 
normalized at each epoch, and equally distributed between the minimum (black) and maximum 
(white) intensities in the shell.}
\label{AzimDistri}
\end{figure}

However, the symmetry of the radio shell not only depends on its circularity, but also on 
the intensity distribution inside it. To study the azimuthal 
intensity distribution in the shell, we computed, for each epoch since 1995, the angular 
distribution of flux density in a ring of radius equal to the radial position of the 
brightness peak.
For every epoch, we used a convolving beam with FWHM equal to 0.5 times the shell 
radius. In Fig. \ref{AzimDistri} we show the time evolution of the angular intensity 
distribution in the SN\,1993J shell, obtained from a linear interpolation between epochs. In 
the cases of epochs observed less than 50 days apart, we selected only one for the interpolation, 
that of highest dynamic range\footnote{The images of all these epochs can be downloaded 
from http://www.uv.es/radioast/aanda2010/sn93j-images.html}. The minimum-to-maximum 
intensity ratio at each epoch typically ranges between 0.7 and 0.9, indicating that the 
shell emission is homogeneous to a level of $\sim$80\%.

From Fig. \ref{AzimDistri} we arrive at a clear conclusion: there are some 
regions where the shell is clearly brighter (i.~e., has hot spots), and these 
regions persist in time for periods of the order of a thousand days. Unfortunately, 
the dynamic range of the images is not high enough to ensure a single interpretation of 
the azimuthal evolution of the radio shell.
The first hot spot is located in the west (i.~e., position angle of 270 degrees in Fig. 
\ref{AzimDistri}) and is present from the beginning of the interpolation up to day 
$\sim$1600 after explosion. There is another, less clear, hot spot present during 
approximately the same time range, but located in the east (i.~e., position angle of 90 
degrees). This hot spot seems to decompose in several parts at some epochs, which
drift towards north and/or south. Beginning on day $\sim$1600 after explosion, 
this second hot spot seems to be finally decomposed into two hot spots, one 
towards the south (reaching a position angle 
$\sim$160 degrees) and the other towards the north (reaching a position angle 
$\sim$0 degrees). These two new hot spots persist in time beyond day 3500 after 
explosion. From that epoch onwards, the dynamic range of our images is too low to reach 
to any robust conclusion about the evolution of the angular brightness pattern. It is 
remarkable that the first hot spot located at a position angle of 270 degrees disappears 
more or less at the same time as the other hot spot evolves into two hot spots which 
shift to their final positions at 160 and 0 degrees. Another possible
interpretation of Fig. \ref{AzimDistri} is that the hot spot 
located at 270 degrees could shift to 160 degrees on day $\sim$1600, and the one located 
at 90 degrees could shift to $\sim$0 degrees at roughly the same time. These 
interpretations of Fig. \ref{AzimDistri} differ from the evolution reported in 
Bietenholz, Bartel \& Rupen (\cite{Bietenholz2003}). According to these authors, 
an additional hot spot should be visible at 180 degrees from day 774 to day 1258, which
is not seen in Fig. \ref{AzimDistri}. However, as we noticed above, for some epochs like 
that of day 1177, the wide hot spot at 90 degrees seems to decompose in several parts, and 
one of the parts shifts close to the south, thus making the figure 
compatible with the reported hot spot in Bietenholz et al. (\cite{Bietenholz2003}). 
According to Bietenholz et al. (\cite{Bietenholz2003}), another hot spot should also 
develop at 270 degrees, beginning on day 2080. A close look to Fig. \ref{AzimDistri} 
reveals that there are indeed small levels of over-emission 
between $\sim200$ and $\sim360$ degrees during practically the whole expansion. The additional 
hot spot reported in Bietenholz et al. (\cite{Bietenholz2003}) could be related to some of the
hot spots shown in our Fig. \ref{AzimDistri}. We must notice, however,
that Fig. \ref{AzimDistri} shows the intensity distribution along a circular ring of 
radius equal to that of the distance of the peak flux density to the shell center. 
Therefore, if a hotspot were at a distance from the shell center different 
from that of the peak, the azimuthal distribution shown in Fig. \ref{AzimDistri} 
would underestimate the intensity of this hot spot (and could also slightly missplace it), 
since the ring used in the azimuthal sampling would not cover the peak of such a hot spot. 

The azimuthal intensity distribution in the shell can either come from inhomogeneities 
in the distribution of the magnetic-field energy density (being higher at the regions with 
higher flux density) or from inhomogeneities in the CSM density distribution (being 
higher at the regions of higher flux density). However, we must take into account that the 
incomplete sampling of the {\em uv}-plane by the interferometers can also partially 
affect the recovered azimuthal intensity distributions, as explained in 
Heywood et al. (\cite{Heywood2009}). Nevertheless, the clear 
systematic evolution of the features and the fact that they persist in the images 
even if we remove several antennas from the data (the hot spots are clearly encoded 
in the phase closures, which are independent of the antenna gains), give us confidence 
in the reliability of the results shown in Fig. \ref{AzimDistri}.

In the case of magnetic-field inhomogeneities inside the shell, a variation 
in the angular distribution of the magnetic-field energy density would be due to 
anisotropies in the ejecta distribution, given that the magnetic fields are presumably 
amplified by nonlinear effects produced in the magneto-hydrodynamic interaction between 
the ejecta and the shocked CSM (see Chevalier \cite{Chevalier1982a}). In the case of 
CSM inhomogeneities, an anisotropic pre-supernova stellar wind 
could help explain the hot spots reported here. 

It must be noticed that the separation between the two hot spots that persist until day 
$\sim$1600 is approximately 180 degrees, which is approximately the same final angular 
separation of the two hot spots that develop after that day. Moreover, the angles between 
the location of the first hot spot (at 270 degrees) and the other hot spots that develop 
after day $\sim$1600 are around 90 degrees. These peculiar angular separations could 
give clues to an interpretation of the hot spots as caused by an anisotropic 
pre-supernova stellar wind along the rotation axis and/or the equatorial plane of the 
progenitor. However, we do not consider the quality of Fig. \ref{AzimDistri} good 
enough to propose any such specific model.

\section{Summary}

We have re-analyzed all the available VLBI data of supernova SN\,1993J in a homogeneous
and self-consistent way. 
We find that the location of the supernova shell center, taking the location 
of the phase calibrator source (M81*) as a reference position, reflects the shift 
in the peak of emission of the 
calibrator with frequency (see also Bietenholz et al. \cite{Bietenholz2004}). 
We find no evidence of proper motion by the shell center at any frequency.

We have obtained expansion curves using two approaches in the data analysis: a novel 
method to estimate the shell size on the sky plane (the CPM, see 
Mart\'i-Vidal \cite{MartiVidal2008} and Marcaide et al. \cite{Marcaide2009}) and model fitting 
to the visibilities. The expansion curves obtained with these approaches can be modeled with 
the same expansion law, and the fitted parameters obtained are compatible. As previously 
found by Marcaide et al. (\cite{Marcaide2009}), the expansion curve differs for different 
observing frequencies. For data at 1.7\,GHz, we can model the expansion with only one expansion
index, $m_3 = 0.87 \pm 0.02$. For data at the other, higher, frequencies, two expansion indices
are needed, $m_1 = 0.933\pm0.010$ and $m_2 = 0.796 \pm 0.005$, which describe the expansion 
in two different regimes separated by a break time $t_{br}=390 \pm 30$ days after explosion.
These estimates are slightly different from those given by Marcaide et 
al. (\cite{Marcaide2009}) and are heavily influenced by having used of early data obtained 
by Bartel's group.

We estimated the shell width at different frequencies and its possible evolution with time, as well
as the opacity to the radio emission by the ejecta. The mean fractional shell width found using
data at all frequencies is $0.31\pm0.02$, compatible with the results previously reported in 
Marcaide et al. (\cite{Marcaide1997}) and Marcaide et al. (\cite{Marcaide2009}), but wider 
than the results reported by Bartel et al. (\cite{Bartel2000}) and Bietenholz et al. 
(\cite{Bietenholz2003}). Our study is not conclusive about any spectral dependence of the shell 
width or any time evolution. With regard to the ejecta absorption, 
we find that our best-quality data can be fitted with a partially-absorbed shell with 
$(80\pm8)$\% absorption. We find evidence of a radial gradient in the distribution of 
the spectral index, indicating that the ejecta opacity may be different
(higher) at 1.7\,GHz, compared to that at higher frequencies, as pointed out in Marcaide et 
al. (\cite{Marcaide2009}). A 
more detailed study of the effect of a frequency-dependent ejecta opacity in the expansion
curve and in the radio light curves is reported in Paper II.

We studied the morphological evolution of the radio shell beyond self-similarity. The 
inhomogeneities (hot spots) found around the azimuthal structure of the shell images are 
typically on the order of 20\% of the mean flux density per unit beam in the shell. The hot 
spots found in the shell persist for times of the order of 1000 days and the angular separations 
between them usually take singular values ($\sim90$ and/or $\sim180$ degrees). This could be 
interpreted as the result of an anisotropic pre-supernova stellar wind along the rotation axis and/or 
the equatorial plane of the precursor star. However, the quality of the data is 
not good enough to reach any robust conclusion.

Shell sizes at very late epochs (from day 3500$-$4000 onwards) are systematically and 
progressively smaller than predicted with the expansion model. This effect may be related to 
the exponential-like time decay of the supernova flux density at late epochs reported 
in Weiler et al. (\cite{Weiler2007}).

\acknowledgements{
We thank the anonymous referee for his/her corrections, detailed analysis and good suggestions 
for improving the paper. IMV is a fellow of the Alexander von Humboldt Foundation in Germany.
The National Radio Astronomy Observatory is a facility of the
National Science Foundation operated under cooperative agreement
with Associated Universities, Inc. The European VLBI Network is a
joint facility of European,Chinese, South African and other radio
astronomy institutes funded by their national research councils.
Partial support from Spanish grant AYA~2005-08561-C03,
AYA~2006-14986-C02, and Prometeo 2009/104 is acknowledged. AA and MAPT 
also acknowledge support by the Consejer\'{\i}a de Innovaci\'on, 
Ciencia y Empresa of Junta de Andaluc\'{\i}a through grants 
FQM-1747 and TIC-126.}

\onecolumn


\longtab{1}{
\begin{longtable}{ l c c | c c c c r }
\caption{VLBI observations of SN\,1993J.}\\

\hline\hline
Date & Age\footnotemark[1] & PI\footnotemark[2] & Freq. & $R_{\mathrm{SC}}$\footnotemark[3] & 
$R_{\mathrm{PR}}$\footnotemark[4] & $R_{\mathrm{MF}}$\footnotemark[5] & Dyn. Range \footnotemark[6] \\
(dd/mm/yy) & (days) & & (GHz) & (mas) & (mas) & (mas) &  \\
\hline

\endfirsthead

\caption{continued.}\\

\hline\hline
Date & Age\footnotemark[1] & PI\footnotemark[2] & Freq. & $R_{\mathrm{SC}}$\footnotemark[3] & 
$R_{\mathrm{PR}}$\footnotemark[4] & $R_{\mathrm{MF}}$\footnotemark[5] & Dyn. Range \footnotemark[6] \\
(dd/mm/yy) & (days) & & (GHz) & (mas) & (mas) & (mas) & \\
\hline
\endhead

\hline
\endfoot

27/04/93\footnotemark[7]
          &  30  & B  &  22 & 0.094\,$\pm$\,0.004 & 0.094\,$\pm$\,0.004 &0.089\,$\pm$\,0.008& ... \\
17/05/93\footnotemark[7]  
          &  50  & B  &  22 & 0.154\,$\pm$\,0.005 & 0.154\,$\pm$\,0.005 &0.146\,$\pm$\,0.010& ... \\*
          &      &    &  15 & 0.162\,$\pm$\,0.018 & 0.162\,$\pm$\,0.018 & 0.15\,$\pm$\,0.03 & ... \\* 
          &      &    & 8.4 & 0.134\,$\pm$\,0.005 & 0.134\,$\pm$\,0.005 &0.127\,$\pm$\,0.009& ... \\ 
27/06/93\footnotemark[7]  
          &  91  & B  &  15 & 0.254\,$\pm$\,0.008 & 0.254\,$\pm$\,0.008 &0.241\,$\pm$\,0.015& ... \\*
          &      &    & 8.4 & 0.264\,$\pm$\,0.006 & 0.264\,$\pm$\,0.006 &0.250\,$\pm$\,0.010& ... \\ 
04/08/93\footnotemark[7]
          & 129  & B  &  15 &  0.38\,$\pm$\,0.03  &  0.38\,$\pm$\,0.03  & 0.36\,$\pm$\,0.06 & ... \\* 
          &      &    & 8.4 & 0.347\,$\pm$\,0.008 & 0.347\,$\pm$\,0.009 &0.329\,$\pm$\,0.015& ... \\ 
19/09/93\footnotemark[7]
          & 175  & B  &  15 & 0.461\,$\pm$\,0.008 & 0.461\,$\pm$\,0.008 &0.437\,$\pm$\,0.015& ... \\* 
          &      &    & 8.4 & 0.448\,$\pm$\,0.008 & 0.448\,$\pm$\,0.009 &0.425\,$\pm$\,0.015& ... \\ 
26/09/93\footnotemark[8] 
          & 182  & M  & 8.4 & 0.517\,$\pm$\,0.017 & 0.517\,$\pm$\,0.018 & 0.49\,$\pm$\,0.03 & 307.4 \\
06/11/93\footnotemark[7] 
          & 223  & B  & 15  & 0.565\,$\pm$\,0.015 & 0.565\,$\pm$\,0.015 & 0.54\,$\pm$\,0.03 & ... \\*
          &      &    & 8.4 & 0.561\,$\pm$\,0.008 & 0.561\,$\pm$\,0.009 &0.531\,$\pm$\,0.015& ... \\* 
          &      &    & 5.0 & 0.565\,$\pm$\,0.017 & 0.565\,$\pm$\,0.017 & 0.54\,$\pm$\,0.03 & ... \\ 
22/11/93\footnotemark[8] 
          & 239  & M  & 8.4 &  0.67\,$\pm$\,0.06  &  0.67\,$\pm$\,0.06  & 0.63\,$\pm$\,0.11 & 450.9 \\
17/12/93\footnotemark[7] 
          & 264  & B  &  15 &  0.77\,$\pm$\,0.03  &  0.77\,$\pm$\,0.03  & 0.73\,$\pm$\,0.05 & ... \\*
          &      &    & 8.4 & 0.710\,$\pm$\,0.011 & 0.710\,$\pm$\,0.011 &0.673\,$\pm$\,0.019& ... \\ 
28/01/94\footnotemark[7] 
          & 306  & B  & 15  &  0.73\,$\pm$\,0.04  &  0.73\,$\pm$\,0.04  & 0.69\,$\pm$\,0.07 & ... \\*
          &      &    & 8.4 & 0.793\,$\pm$\,0.014 & 0.793\,$\pm$\,0.014 & 0.75\,$\pm$\,0.02 & ... \\* 
          &      &    & 5.0 &  0.78\,$\pm$\,0.02  &  0.78\,$\pm$\,0.02  & 0.74\,$\pm$\,0.04 & ... \\ 
20/02/94\footnotemark[8] 
          & 329  & M  & 8.4 &  0.87\,$\pm$\,0.07  &  0.87\,$\pm$\,0.07  & 0.82\,$\pm$\,0.12 & 199.3 \\
15/03/94\footnotemark[7] 
          & 352  & B  & 8.4 & 0.862\,$\pm$\,0.014 & 0.862\,$\pm$\,0.014 & 0.82\,$\pm$\,0.02 & ... \\*
          &      &    & 5.0 & 0.863\,$\pm$\,0.014 & 0.863\,$\pm$\,0.014 & 0.82\,$\pm$\,0.02 & ... \\ 
22/04/94\footnotemark[7] 
          & 390  & B  & 8.4 & 0.991\,$\pm$\,0.017 & 0.991\,$\pm$\,0.017 & 0.94\,$\pm$\,0.03 & ... \\*
          &      &    & 5.0 & 0.955\,$\pm$\,0.017 & 0.955\,$\pm$\,0.017 & 0.90\,$\pm$\,0.03 & ... \\* 
          &      &    & 2.3 &  1.00\,$\pm$\,0.05  &  1.00\,$\pm$\,0.05  & 0.95\,$\pm$\,0.09 & ... \\ 
29/05/94\footnotemark[8] 
          & 427  & M  & 8.4 &  1.08\,$\pm$\,0.10  &  1.08\,$\pm$\,0.11  & 1.02\,$\pm$\,0.18 & 123.8 \\
22/06/94\footnotemark[7] 
          & 451  & B  & 8.4 & 1.062\,$\pm$\,0.017 & 1.062\,$\pm$\,0.017 & 1.01\,$\pm$\,0.03 & ... \\*
          &      &    & 5.0 & 1.098\,$\pm$\,0.017 & 1.098\,$\pm$\,0.017 & 1.04\,$\pm$\,0.03 & ... \\ 
30/08/94\footnotemark[7] 
          & 520  & B  & 8.4 & 1.248\,$\pm$\,0.019 &  1.25\,$\pm$\,0.02  & 1.18\,$\pm$\,0.03 & ... \\*
          &      &    & 5.0 & 1.232\,$\pm$\,0.019 &  1.23\,$\pm$\,0.02  & 1.17\,$\pm$\,0.03 & ... \\* 
          &      &    & 2.3 &  1.29\,$\pm$\,0.04  &  1.29\,$\pm$\,0.04  & 1.22\,$\pm$\,0.08 & ... \\ 
20/09/94\footnotemark[8] 
          & 541  & M  & 5.0 &  1.21\,$\pm$\,0.14  &  1.21\,$\pm$\,0.14  & 1.15\,$\pm$\,0.24 & 152.4 \\
31/10/94\footnotemark[7] 
          & 582  & B  & 8.4 &  1.35\,$\pm$\,0.02  &  1.35\,$\pm$\,0.02  & 1.28\,$\pm$\,0.04 & ... \\*
          &      &    & 5.0 &  1.35\,$\pm$\,0.02  &  1.35\,$\pm$\,0.02  & 1.27\,$\pm$\,0.04 & ... \\ 
23/12/94\footnotemark[7] 
          & 635  & B  & 8.4 &  1.39\,$\pm$\,0.02  &  1.39\,$\pm$\,0.02  & 1.31\,$\pm$\,0.04 & ... \\*
          &      &    & 5.0 &  1.43\,$\pm$\,0.02  &  1.43\,$\pm$\,0.02  & 1.36\,$\pm$\,0.04 & ... \\* 
          &      &    & 2.3 &  1.48\,$\pm$\,0.05  &  1.48\,$\pm$\,0.05  & 1.40\,$\pm$\,0.09 & ... \\ 
12/02/95  & 686  & B  & 8.4 &  1.52\,$\pm$\,0.08  &  1.53\,$\pm$\,0.08  & 1.41\,$\pm$\,0.04 & 42.7 \\
23/02/95  & 697  & M  & 5.0 &  1.48\,$\pm$\,0.04  &  1.48\,$\pm$\,0.04  & 1.38\,$\pm$\,0.11 & 140.7 \\
11/05/95  & 774  & B  & 8.4 &  1.60\,$\pm$\,0.20  &  1.64\,$\pm$\,0.21  & 1.68\,$\pm$\,0.12 & 36.9 \\*
          &      &    & 5.0 &  1.68\,$\pm$\,0.10  &  1.68\,$\pm$\,0.11  & 1.56\,$\pm$\,0.04 & 14.2 \\ 
11/05/95  & 774  & M  & 5.0 &  1.66\,$\pm$\,0.07  &  1.71\,$\pm$\,0.08  & 1.51\,$\pm$\,0.80 & 194.6 \\
18/08/95  & 873  & B  & 8.4 &  1.73\,$\pm$\,0.16  &  1.73\,$\pm$\,0.16  & 1.77\,$\pm$\,0.11 & 23.5 \\*
          &      &    & 5.0 &  1.85\,$\pm$\,0.07  &  1.81\,$\pm$\,0.07  & 1.51\,$\pm$\,0.30 & 44.5 \\ 
01/10/95  & 917  & M  & 5.0 &  1.92\,$\pm$\,0.07  &  1.92\,$\pm$\,0.08  & 1.92\,$\pm$\,0.05 & 168.7 \\
19/12/95  & 996  & B  & 8.4 &  2.11\,$\pm$\,0.06  &  2.11\,$\pm$\,0.06  & 2.12\,$\pm$\,0.11 & 35.2 \\*
          &      &    & 5.0 &  2.15\,$\pm$\,0.11  &  2.15\,$\pm$\,0.12  & 2.13\,$\pm$\,0.07 & 71.0 \\
          &      &    & 2.3 &  2.23\,$\pm$\,0.05  &  2.23\,$\pm$\,0.05  & 2.11\,$\pm$\,0.08 & 116.9 \\
28/03/96  & 1096 & M  & 5.0 &  2.21\,$\pm$\,0.05  &  2.21\,$\pm$\,0.06  & 2.14\,$\pm$\,0.11 & 63.0 \\
08/04/96  & 1107 & B  & 8.4 &  2.41\,$\pm$\,0.18  &  2.34\,$\pm$\,0.19  & 2.14\,$\pm$\,0.15 & 52.3 \\*
          &      &    & 5.0 &  2.10\,$\pm$\,0.20  &  2.02\,$\pm$\,0.21  & 1.98\,$\pm$\,0.12 & 56.0 \\ 
17/06/96  & 1177 & M  & 5.0 &  2.31\,$\pm$\,0.07  &  2.31\,$\pm$\,0.07  & 2.23\,$\pm$\,0.10 & 30.6 \\
01/09/96  & 1253 & B  & 8.4 &  2.50\,$\pm$\,0.10  &  2.53\,$\pm$\,0.10  & 2.29\,$\pm$\,0.06 & 30.7 \\*
          &      &    & 5.0 &  2.51\,$\pm$\,0.08  &  2.50\,$\pm$\,0.08  & 2.33\,$\pm$\,0.04 & 74.5 \\ 
22/10/96  & 1304 & M  & 5.0 &  2.61\,$\pm$\,0.04  &  2.61\,$\pm$\,0.04  & 2.47\,$\pm$\,0.06 & 120.9 \\
13/12/96  & 1356 & B  & 8.4 &  2.65\,$\pm$\,0.06  &  2.66\,$\pm$\,0.06  & 2.52\,$\pm$\,0.07 & 20.6 \\*
          &      &    & 5.0 &  2.62\,$\pm$\,0.11  &  2.63\,$\pm$\,0.12  & 2.59\,$\pm$\,0.11 & 62.8 \\* 
          &      &    & 2.3 &  2.52\,$\pm$\,0.18  &  2.52\,$\pm$\,0.19  & 2.44\,$\pm$\,0.10 & 165.1 \\ 
25/02/97  & 1430 & M  & 5.0 &  2.81\,$\pm$\,0.09  &  2.81\,$\pm$\,0.10  & 2.51\,$\pm$\,0.10 & 79.4 \\
07/06/97  & 1532 & B  & 8.4 &  2.90\,$\pm$\,0.13  &  2.90\,$\pm$\,0.14  & 2.73\,$\pm$\,0.11 & 19.1 \\*
          &      &    & 5.0 &  2.86\,$\pm$\,0.06  &  2.84\,$\pm$\,0.07  & 2.70\,$\pm$\,0.43 & 21.6 \\ 
21/09/97  & 1638 & M  & 5.0 &  3.09\,$\pm$\,0.05  &  3.09\,$\pm$\,0.06  & 2.94\,$\pm$\,0.02 & 101.9 \\
15/11/97  & 1693 & B  & 8.4 &  3.14\,$\pm$\,0.07  &  3.15\,$\pm$\,0.12  & 2.91\,$\pm$\,0.15 & 31.2 \\*
          &      &    & 5.0 &  3.13\,$\pm$\,0.15  &  3.13\,$\pm$\,0.16  & 3.05\,$\pm$\,0.13 & 84.0 \\* 
          &      &    & 2.3 &  3.17\,$\pm$\,0.50  &  3.17\,$\pm$\,0.57  & 3.03\,$\pm$\,0.16 & 113.6 \\ 
18/02/98  & 1788 & M  & 5.0 &  3.37\,$\pm$\,0.05  &  3.39\,$\pm$\,0.05  & 3.21\,$\pm$\,0.05 & 86.6 \\
30/05/98  & 1889 & M  & 5.0 &  3.48\,$\pm$\,0.06  &  3.48\,$\pm$\,0.07  & 3.33\,$\pm$\,0.09 & 112.6 \\
03/06/98  & 1893 & B  & 8.4 &  3.46\,$\pm$\,0.07  &  3.37\,$\pm$\,0.08  & 3.33\,$\pm$\,0.14 & 35.4 \\*
          &      &    & 5.0 &  3.43\,$\pm$\,0.09  &  3.45\,$\pm$\,0.09  & 3.18\,$\pm$\,0.13 & 26.9 \\ 
20/11/98  & 2064 & B  & 5.0 &  3.73\,$\pm$\,0.16  &  3.75\,$\pm$\,0.17  & 3.63\,$\pm$\,0.06 & 26.0 \\
23/11/98  & 2066 & M  & 5.0 &  3.74\,$\pm$\,0.07  &  3.74\,$\pm$\,0.08  & 3.50\,$\pm$\,0.05 & 94.7 \\
30/11/98  & 2073 & M  & 1.7 &  3.78\,$\pm$\,0.07  &  3.91\,$\pm$\,0.08  & 3.58\,$\pm$\,0.15 & 150.5 \\
07/12/98  & 2080 & B  & 8.4 &  3.74\,$\pm$\,0.09  &  3.75\,$\pm$\,0.09  & 3.47\,$\pm$\,0.14 & 30.0 \\*
          &      &    & 2.3 &  3.77\,$\pm$\,0.07  &  3.96\,$\pm$\,0.08  & 3.57\,$\pm$\,0.16 & 62.0 \\ 
06/06/99  & 2261 & B  & 1.7 &  4.11\,$\pm$\,0.07  &  4.17\,$\pm$\,0.08  & 3.82\,$\pm$\,0.05 & 60.1 \\
10/06/99  & 2265 & M  & 5.0 &  4.07\,$\pm$\,0.08  &  4.10\,$\pm$\,0.08  & 3.81\,$\pm$\,0.04 & 61.9 \\
16/06/99  & 2271 & B  & 5.0 &  4.05\,$\pm$\,0.35  &  4.02\,$\pm$\,0.36  & 3.71\,$\pm$\,0.16 & 23.3 \\
22/09/99  & 2369 & M  & 5.0 &  4.12\,$\pm$\,0.07  &  4.12\,$\pm$\,0.08  & 3.93\,$\pm$\,0.05 & 41.1 \\
28/09/99  & 2376 & B  & 1.7 &  4.28\,$\pm$\,0.07  &  4.43\,$\pm$\,0.08  & 4.13\,$\pm$\,0.09 & 3.8 \\
24/11/99  & 2432 & B  & 5.0 &  4.32\,$\pm$\,0.22  &  4.34\,$\pm$\,0.23  & 4.12\,$\pm$\,0.12 & 21.8 \\
25/02/00  & 2525 & B  & 8.4 &  4.43\,$\pm$\,0.09  &  4.48\,$\pm$\,0.10  & 4.12\,$\pm$\,0.32 & 26.9 \\
06/06/00  & 2627 & M  & 5.0 &  4.52\,$\pm$\,0.10  &  4.38\,$\pm$\,0.11  & 4.36\,$\pm$\,0.06 & 40.2 \\
13/11/00  & 2787 & B  & 8.4 &  4.77\,$\pm$\,0.13  &  4.65\,$\pm$\,0.13  & 4.90\,$\pm$\,0.29 & 14.5 \\*
          &      &    & 2.3 &  4.85\,$\pm$\,0.40  &  4.70\,$\pm$\,0.39  & 4.81\,$\pm$\,0.25 & 19.5 \\ 
20/11/00  & 2794 & M  & 1.7 &  5.01\,$\pm$\,0.13  &  5.07\,$\pm$\,0.13  & 4.75\,$\pm$\,0.05 & 219.7 \\
24/11/00  & 2798 & M  & 5.0 &  4.76\,$\pm$\,0.18  &  4.79\,$\pm$\,0.18  & 4.76\,$\pm$\,0.06 & 56.71 \\
14/02/01  & 2880 & M  & 5.0 &  4.94\,$\pm$\,0.12  &  4.95\,$\pm$\,0.12  & 4.65\,$\pm$\,0.07 & 116.2 \\
10/06/01  & 2996 & B  & 5.0 &  5.00\,$\pm$\,0.14  &  4.99\,$\pm$\,0.14  & 4.90\,$\pm$\,0.09 & 32.1 \\
18/11/01  & 3157 & M  & 5.0 &  5.21\,$\pm$\,0.18  &  5.16\,$\pm$\,0.19  & 4.28\,$\pm$\,0.60 & 38.7 \\
26/11/01  & 3164 & B  & 1.7 &  5.59\,$\pm$\,0.17  &  5.52\,$\pm$\,0.18  & 5.17\,$\pm$\,0.12 & 98.1 \\
24/05/02  & 3344 & B  & 5.0 &  5.49\,$\pm$\,0.18  &  5.48\,$\pm$\,0.19  & 5.03\,$\pm$\,0.06 & 13.2 \\
07/11/02  & 3511 & M  & 5.0 &  5.77\,$\pm$\,0.19  &  5.76\,$\pm$\,0.19  & 5.57\,$\pm$\,0.40 & 18.2 \\
17/11/02  & 3521 & M  & 1.7 &  6.15\,$\pm$\,0.18  &  6.06\,$\pm$\,0.19  & 5.85\,$\pm$\,0.10 & 78.5 \\
01/06/03  & 3717 & B  & 5.0 &  6.04\,$\pm$\,0.23  &  5.79\,$\pm$\,0.23  & 5.26\,$\pm$\,0.29 & 19.6 \\
29/10/03  & 3867 & M  & 5.0 &  6.34\,$\pm$\,0.18  &  6.34\,$\pm$\,0.20  & 5.88\,$\pm$\,0.22 & 21.9 \\
25/05/04  & 4076 & B  & 5.0 &  6.45\,$\pm$\,0.31  &  6.13\,$\pm$\,0.32  & 5.74\,$\pm$\,0.30 & 15.3 \\
06/06/04  & 4088 & B  & 1.7 &  6.72\,$\pm$\,0.21  &  6.58\,$\pm$\,0.24  & 6.49\,$\pm$\,0.14 & 35.4 \\
31/10/04  & 4235 & M  & 5.0 &  6.33\,$\pm$\,0.24  &  6.56\,$\pm$\,0.24  &...\footnotemark[9]& 19.8 \\
11/06/05  & 4458 & B  & 5.0 &  6.55\,$\pm$\,0.26  &  6.89\,$\pm$\,0.27  & 5.90\,$\pm$\,0.61 & 18.7 \\
22/10/05  & 4591 & M  & 5.0 &  6.64\,$\pm$\,0.29  &  6.64\,$\pm$\,0.29  & 6.10\,$\pm$\,0.57 & 27.3 \\
06/11/05  & 4606 & M  & 1.7 &  7.00\,$\pm$\,0.30  &  6.87\,$\pm$\,0.27  & 6.58\,$\pm$\,0.11 & 45.9 \\ 

\footnotetext[1]{Assumed explosion date: ~~~ 28 March 1993.}
\footnotetext[2]{Principal Investigator:~~~ M = J.M. Marcaide; ~~~ B = N. Bartel / M.F. Bietenholz / M. Rupen.}
\footnotetext[3]{Shell radii determined with the CPM applied to images obtained from self-calibrated 
visibilities (see text).}
\footnotetext[4]{Shell radii determined with the CPM applied to images obtained from phase-referenced 
visibilities (see text).}
\footnotetext[5]{Shell radii determined from model fitting to the visibilities.}
\footnotetext[6]{Dynamic range of the supernova images (i.e., peak flux density in units of the 
root-mean-square of the image background), computed using natural visibility weighting and applying a Gaussian 
taper in Fourier space (see Sect. \ref{Calibracion}).}
\footnotetext[7]{Epochs where the shell radii were taken from Bartel et al. (\cite{Bartel2002}), but applying 
the corresponding biases (see Sect. \ref{ExpansionCPM}) to make them comparable to the radii obtained with the 
CPM and with the radii estimated from our visibility model fitting.}
\footnotetext[8]{Epochs where the ``CPM-like'' radii were adapted from model-fitting results, to avoid image 
over-resolutions if the CPM were applied directly.}
\footnotetext[9]{Unsatisfactory fit (unclear minimum of the $\chi^2$).}

\end{longtable}
\label{SN93JAllObs}
}

\end{document}